\documentclass[osajnl,twocolumn,showpacs,superscriptaddress,10pt]{revtex4-1} 
\usepackage{graphicx}
\usepackage{hyperref}
\usepackage{amssymb,amsmath}
\usepackage{epstopdf}
\begin{document}

\title{Ultra-fast heralded single photon source based on telecom technology }
\author{Lutfi Arif Ngah}
\author{Olivier Alibart}
\author{Laurent Labont\'e}
\author{Virginia D'Auria}\email{Corresponding author: virginia.dauria@unice.fr}
\author{S\'ebastien Tanzilli}
\affiliation{Laboratoire de Physique de la Mati\`ere Condens\'ee, CNRS UMR 7336, Universit\'e Nice Sophia Antipolis, Parc Valrose, 06108 Nice Cedex 2, France}

\begin{abstract} 

The realization of an ultra-fast source of heralded single photons emitted at the wavelength of 1540 nm is reported. The presented strategy is based on state-of-the-art telecom technology, combined with off-the-shelf fiber components and waveguide non-linear stages pumped by a 10 GHz repetition rate laser. The single photons are heralded at a rate as high as 2.1 MHz with a heralding efficiency of 42\%. Single photon character of the source is inferred by measuring the second-order autocorrelation function. For the highest heralding rate, a value as low as 0.023 is found. This not only proves negligible multi-photon contributions but also represents the best measured value reported to date for heralding rates in the MHz regime. These prime performances, associated with a device-like configuration, are key ingredients for both fast and secure quantum communication protocols.

\end{abstract}

\maketitle 

The reliable generation of single photon states is crucial for a wide variety of quantum optical technologies, ranging from quantum computation and communication \cite {Polyakov2011, Gisin2007} to quantum metrology and detector calibration \cite {Brida2011, Migdall2009}. As an example, the use of \emph{single} photon states is essential in quantum key distribution (QKD) protocols, where the unintended presence of more than one photon per time window can be exploited by an eavesdropper to extract part of the information \cite{Scarani2009}.

\begin{figure*}
\begin{center}
\includegraphics[width=2\columnwidth]{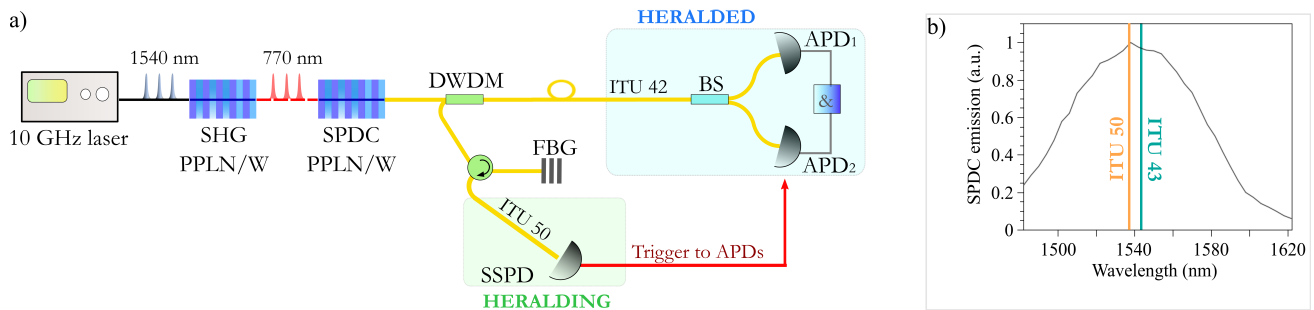}
\caption{a) Schematic experimental setup and b) SPDC emission. The telecom laser with a repetition rate of 10 GHz is frequency converted via second harmonic generation (SHG) and then used to pump the spontaneous parametric down conversion (SPDC). Both the processes take place in home-made periodically poled lithium niobate waveguides (PPLN/Ws). At the output of SPDC, signal photon at 1537.40 (telecom ITU channel 50) and idler photons at 1543.73 (ITU 42) are selected within the SPDC spectrum and deterministically separated thanks to the combination of dense wavelength demultiplexer (DWDM) and a narrow-band Bragg filter (FBG). Signal photons are directed toward the heralding path and detected by a super conducting single photon detector (SSPD). Heralded idler photons are analyzed with an Hanbury Brown-Twiss (HBT) setup made of a beam splitter (BS) and two avalanche photodiodes ($APD_1$ and $APD_2$). The detection in the HBT is triggered by the output signal of the SSPD.}
\label{setup}
\end{center}
\end{figure*}

Ideal sources should be able to emit indistinguishable single photons in a deterministic way, at an arbitrarily high  repetition rate and with zero probability of multi-photon emissions \cite{Polyakov2011}. In particular, the request of ultra-fast photon sources is mandatory to speed up data exchanges in quantum communication protocols. In anticipation to such optimal cases, a pertinent alternative is represented by heralded single photon sources (HSPS), where the detection of one photon of two simultaneously generated is used to herald the emission time of the second one \cite{Polyakov2011, reviewSeb}. In such schemes, the produced single photons rate is proportional to the detected heralding photon one, $R_H$, and to the heralding efficiency, $P_1$, namely, the probability of observing one heralded photon per heralding event. We note that, in experiments,  the value of  $P_1$ is essentially determined by optical losses \cite{reviewSeb}.

In the original and most common implementations, pairs of simultaneous photons are generated in non-linear crystals via spontaneous parametric down conversion (SPDC) of a pump beam \cite{reviewSeb}. In particular, an accurate choice of the phase matching can lead to the production of photons at telecom wavelength, as required for long distance transmission in optical fibers \cite{Alibart2005, Aboussouan2010}. SPDC being a probabilistic process, a way to obtain high photon rates is to increase the probability of generating the photon pairs as well as the photon transmission after the SPDC crystal. Accordingly, in the last years, many papers have been focusing on the realization of bright SPDC sources \cite{Oxford, Silberhorn2013} and much effort has been made towards optimizing paired photon collection, separation and propagation \cite{Pomarico2012, Migdall2013, Zeilinger2013, Thew2013}.

 Most of these experiments exploit solid-state lasers to pump the SPDC crystal. Such lasers, however, are able to produce light pulses at repetition rates limited to a few hundreds of MHz. At the same time, the HSPS theory shows that in order to guarantee high quality heralded single photons with negligible two (or more) photons contributions, the mean number of photon pairs produced per pump pulse, $\langle n \rangle$, must be kept well below 0.1 \cite{Scarani2004, Giapponesi2014}. As a matter of fact, in previous implementations, the combination of moderate SPDC pump repetition rates with the requirement of low $\langle n \rangle$ fundamentally limited the speed of the reported HSPSs  \cite{Pomarico2012}.\\
Speeding up the available rate of heralded photons can be achieved by spatial multiplexing of the pump beam over many simultaneous SPDC stages \cite{Migdall2002, Meany2014}. Conversely, in this paper, we make use of state-of-the-art telecom technology and non-linear optics to time-multiplex the pump beam of a single SPDC process. By doing so, we demonstrate an efficient and ultra-fast HSPS, able to generate with $P_{1}\approx 42\%$, high quality telecom single-photons at a measured $R_H$=2.1 MHz. In our scheme, the SPDC crystal is pumped with a pulsed telecom laser working at the ultra-fast repetition rate of 10 GHz. The laser is first frequency doubled via second harmonic generation (SHG) so as to fulfill the SPDC phase matching condition for generating both heralding (signal) and heralded (idler) photons in the C-band of telecom wavelengths. As discussed in \cite{Giapponesi2014} and \cite{Yamamoto2007, Broome2011,Morris2014}, the high-repetition rate laser allows generating photon pairs at a high rate while maintaining $\langle n \rangle$ as low as required for negligible two-photon events. In our work, we provide an experimental proof of this statement, by actually measuring the second order autocorrelation function $g^{(2)}(0)$ of our heralded photons. We also show that the obtained results are the best ever reported for $R_H$ in the MHz regime.

Our experimental setup is schematically depicted in Figure \ref{setup}-a. An ultra-fast telecom laser (Pritel UOC 100) delivers ps-optical pulses at 1540.56 nm (telecom ITU channel 46) at the repetition rate of 10 GHz. The laser output is sent to a home-made periodically poled lithium niobate waveguide (PPLN/W) to be converted via SHG into light pulses at 770.28 nm (spectral FWHM=0.40 nm) and then used to pump the SPDC stage in a second home-made PPLN/W. Our measured SHG conversion efficiency is $\sim$ 20\%. The SPDC waveguide produces, via type-0 quasi-phase matching, pairs of frequency degenerate photons centered around the wavelength of 1540.56 nm within a spectral bandwidth of 80 nm (see Figure \ref{setup}-b). The output light is collected by a butt-coupled optical fiber (SMF 28) and directed toward a dense wavelength demultiplexer (DWDM) followed by a narrow-band Bragg filter (FBG). This filter combination allows picking up, within the SPDC spectrum, signal and idler photons slightly away from perfect degeneracy and thus deterministically separable (see Figure \ref{setup}-b). The entire setup downstream the SPDC operation is made of standard telecom components with overall losses amounting to 2.5 dB on the signal path and to 1.9 dB on the idler one. After the spectral filters and upon correction for imperfect component transmission, we measure for our SPDC a brightness of about 2.5 $\cdot 10^5$ pairs/mW/s/GHz.

The FBG selects signal photons at 1537.40 nm (ITU channel 50) within a bandwidth of 250 pm (i.e. 25 GHz). These are directed toward the heralding photon path where they are eventually detected by a super-conducting single photon detector (SSPD, SCONTEL model LTD 24/30-008) able to handle the MHz-counting regime. Our detector exhibits a quantum detection efficiency $\eta_D \simeq$ 0.17, dark count rate of $\sim$ 100 Hz and a measured time-jitter $\lesssim$ 57 ps. We will designate the SSPD rate as $R_H$. Beside $\eta_D$, in the limit of negligible detector dark counts, the heralding rate depends on the laser repetition rate $f$ and on the photon pair generation probability:
\begin{equation}
  \label{eq:RH}
  R_H=f \cdot \langle n \rangle \cdot \gamma \cdot T_H \cdot \eta_D
  \,,
\end{equation}
where, in this case, $\langle n \rangle$ denotes the mean number of photon emitted in the heralding photon bandwidth per pump pulse. Moreover, $\gamma$ represents the coupling efficiency between the SPDC waveguide and the collection fiber at its output, and $T_H$ stands for the transmission from the collection fiber to the SSPD.

Heralded idler photons at  1543.73 nm (ITU channel 42) with a bandwidth of 1.1 nm (i.e. 200 GHz) are transmitted by the DWDM toward a fibered Hanbury-Brown-Twiss type setup (HBT) \cite{HBT} to be characterized in terms of $P_1$ and the autocorrelation function, $g^{(2)}(0)$. The heralded photon bandwidth has been chosen so as to adequately collect the idler photons who are energy-matched with the heralding ones given the spectral width of the SPCD pump. The HBT detectors are InGaAs avalanche photodiodes (respectively $APD_1$, idQuantique 210 and $APD_2$, id Quantique 201) exhibiting each a detection efficiency $\eta_i$= 0.25 (i=1,2) and a time jitter of $\sim$120 ps.  $APD_1$ is triggered by the detection signal from the SSPD, while $APD_2$ is triggered by $S_1$, the detection signal of $APD_1$. Accordingly, the detection rate of $APD_2$, $S_2$, gives the coincidence signal between the two APDs. In order to reduce the contribution of accidental detections of signal and idler photons generated by different pump pulses, at the output of each APD, we use a time-to-amplitude converter (TAC/SCA, Ortec 567) to post-select detection events within a narrow time window after having registered the trigger signal. Ideally, optimal filtering is achieved by a window shorter than the delay between subsequent pump pulses. Experimentally, however, the accuracy on the detection time is essentially down-limited by the detection timing jitters. Accordingly, we set for $APD_1$ and $APD_2$ time windows of $\approx$ 300 ps and $\approx$ 400 ps, respectively. These windows include contributions from 3 and 4 consecutive pump pulses.

\begin{figure}
\includegraphics[width=0.9\columnwidth]{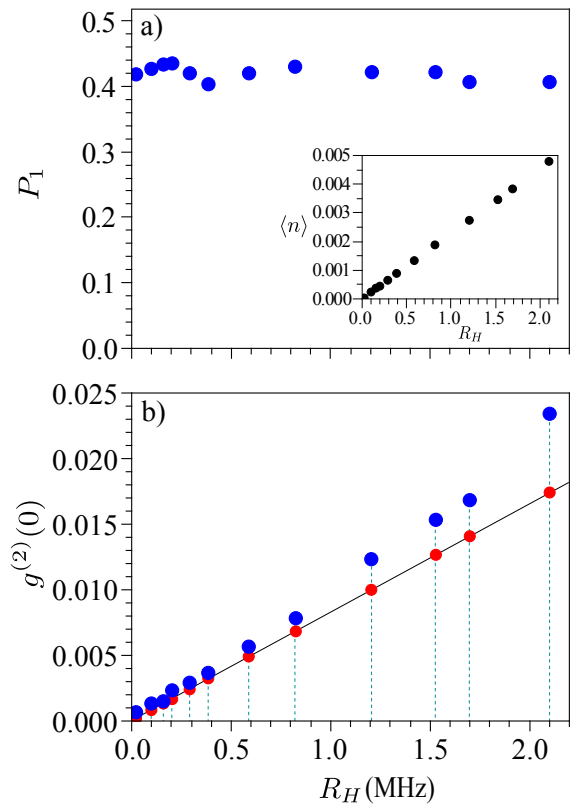}
\caption{Experimental results for the heralding efficiency $P_1$ (a) and for the heralded photon autocorrelation function (b) as functions of the rate of detected heralding photons, $R_H$. The inset of fig.a shows the mean number of produced pairs in the heralding photon bandwidth per pump pulse vs $R_H$. In the fig.b, blue dots represent experimental data and red dots theoretically expected ones in the single mode theory (see Ref. \cite{Sekatski2012}, the line is a guide to the eye).}
\label{Dati}
\end{figure}

As detailed in Ref. \cite{Alibart2005}, the values of $P_1$ and $g^{(2)}(0)$ can be expressed in terms of the conditional detection rates, $S_1$ and $S_2$, by taking into account the non-unitary detection efficiency of the APDs as well as their dark count rates. In the limit of negligible detector dark counts and of $S_2 \ll S_1 $ we obtain:
\begin{equation}
  \label{eq:p1g2}
   P_1\simeq\frac{2 S_1}{R_H\cdot\eta_1} \qquad \textrm{and} \qquad   g^{(2)}(0)\simeq\frac{R_H\cdot S_2 \cdot \eta_1}{S_1^2\cdot\eta_2}
   \,,
\end{equation}
where, in the actual data analysis, $R_H$, $S_i$ and $\eta_i$ (i=1,2) take into account the effect of the APDs dead times (10 $\mu$s for each APD) \cite{Zeilinger2013}. The value of the heralding efficiency linearly depends on the transmission of the heralded photon path, T,  and on the coupling efficiency, $\gamma$, as $P_1 \simeq\gamma \cdot T$. As shown in Figure \ref{Dati}-a, in all of our tests, we measure $P_1 \approx$ 0.42, corresponding to an estimated $\gamma$ of 0.60. Coupling efficiencies as high as 0.80 have been recently measured for SPDC in bulk crystals \cite{Thew2013, Zeilinger2013}, however our choice to use a non-linear interaction in optical waveguides presents the advantage of better SPDC efficiency together with a simplified and more stable setup. Besides, an improvement of our coupling efficiency could be obtained by AR coating the facets of our SPDC crystal \cite{Silberhorn2013}.

In order to experimentally prove the ultra-fast operation of our source and the high quality of the produced single photons, we checked the evolution of our heralded states as a function of increasing values of the mean number of photon pairs generated. Variations of $\langle n \rangle$ can be simply obtained by increasing the power of the pump light at 770.28 nm at the input of the SPDC stage. The experimental values of $\langle n \rangle$ are then retrieved by inverting Eq.\,\eqref{eq:RH}  and dividing the corresponding $R_H$ by the laser repetition rate, and the overall collection, transmission and detection efficiency on the heralding photon path \cite{Pomarico2012}. As already stressed, $\langle n \rangle$ calculated in this way coincides with the mean number of heralding photons generated per pump pulse.
 
We made $R_H$ vary from a few tens of kHz up to 2.1 MHz, which corresponds to a maximum $\langle n \rangle \simeq 0.005$ (see inset of Figure \ref{Dati}-a). Such a value is well below the limit of 0.1 demanded for single photon states with negligible multi-photon contributions \cite{Scarani2004}. This is confirmed by our experimental results for the autocorrelation function, $g^{(2)}(0)$. This quantity is ideally 0 for perfect single photon Fock states but increases with the probability of unwanted two (or more)-photon events. Figure \ref{Dati}-b shows the measured $g^{(2)}(0)$ values (in blue) together with the theoretical prediction for a perfect single mode heralded photon (in red) \cite{Sekatski2012}. Our highest experimental $g^{(2)}(0)$ is 0.023, representing, to our knowledge, the best measured value for the autocorrelation function of heralded single photons announced at a rate in the MHz regime. The mismatch between the theory and the experiments observed for $R_H\geq$1MHz can be explained by non-negligible photonic noise due to unwanted time and spectral modes on the heralded photons. As discussed in \cite{Oxford}, by choosing to collect all heralded photons energy-matched with the heralding ones, we optimize the heralding efficiency $P_1$ at the price of a reduced heralded photon spectral purity (see also \cite{Stefano2010}). The number of spurious spectral modes, $N_f$, is approximately given by the ratio between the heralded and heralding photon bandwidths, which is $\sim8$ in our case. This value must be multiplied by the number, $N_t$, of parasite temporal modes which affect our $g^{(2)}(0)$ measurements and that can be estimated as the ratio between the detection window and the laser repetition rate, $\sim4$ in our experiment. Accordingly, we find that in order to keep the mean number of heralded photon below 0.1, the mean number of heralding photon must be as low as $0.1/(N_f\cdot N_t)\approx 0.003$, in agreement with a heralding rate of $\approx$1 MHz.

In table \ref{tabella}, we compare our results with those obtained in other recent implementations of HSPSs. As it can be seen, our results represent the best observed in terms of fast photon generation and low rate of multi-photon emissions. Besides, we recall that the experimental value of $R_H$ is proportional to the quantum efficiency of the heralding photon detector $\eta_D$. In our experimental setup, heralding photons are detected by an old-generation SSPD showing a $\eta_D \simeq$ 0.17. This value is well below the detection efficiency of Si-APDs used in many previous experiments \cite{Silberhorn2013, Pomarico2012, Zeilinger2013, Brida2012} who relied on non-degenerate emission of pairs of photon with heralding photons in the visible and NIR range of wavelengths. At the same time, we note that new generation SSPDs shows now quantum efficiency at telecom wavelength as high as 90\% \cite{Wang2013, Nam2013}. Accordingly, by just replacing our SSPD with an improved one (and with no other fundamental change in the experimental setup), our observed $R_H$ values would be increased by a factor 5 without introducing any degradation to the value of $g^{(2)}(0)$. As already explained, a further improvement would be that of a better coupling efficiency at the SPDC output up to the optimal case $\sim$ 0.80. For a comparison with results from other groups, the last line of table~\ref{tabella} reports expected results for our scheme, when using state-of-the-art SSPD detectors and exploiting optimized waveguide-to-fiber coupling.

\begin{table}[top]
\begin{tabular}{| l | c |c| c| c| c |} \hline
 & $P_1$& $\eta_D$ & $R_H $ & $\langle n \rangle$ & $g^{(2)}(0)$\\
 
   \hline  
Nice & 0.42 &0.17 & 2.1 MHz & 0.005 & 0.023 \\
	Geneva \cite{Pomarico2012}& 0.45 &0.50& 4.4 MHz & 0.1 & $0.18^a$ \\
	Paderborn \cite{Silberhorn2013} &0.60& 0.55& 105 kHz & -& $0.40$\\
	Milan \cite{Brida2012} & 0.13 &0.40 & $\sim$10 kHz$^b$ & -& 0.0050\\
	Vienna \cite{Zeilinger2013} & 0.82 & 0.95 & 6 kHz & - & -\\
	Tokyo \cite{Giapponesi2014} & $<$0.3&0.70&$\sim$150 kHz$^b$&0.00021&-\\
	Nice$^c$ & 0.5 & 0.90 & 15 MHz & 0.005 & $\lesssim$0.020\\
    \hline
\end{tabular}
\caption{Experimental results for different HSPS realizations. For each group, only the values corresponding to the highest measured $R_H$ has been reported. $^a$ theoretically calculated. $^b$ estimated from reported data and $P_1$. $^c$ expected values.}
\label{tabella}
\end{table}

In conclusion, we have employed advanced telecom technologies combined with non-linear optics to realize an ultra-fast HSPS able to provide high-quality single photons at telecom wavelength. In order to prove the high quality of our heralded single photons we measured their autocorrelation function, obtaining heralding rate up to MHz with $\langle n \rangle < 0.1$ and  $g^{(2)}(0) \leq $ 0.023. These results prove that our source is a valuable candidate for many quantum optics and quantum communication schemes where the requirement of fast data exchange and negligible-multiple photon emission are mandatory.

\begin{acknowledgements}

  The authors acknowledge financial support from the Agence Nationale de la Recherche (ANR) for the $\emph{Conneqt}$ (ANR-2011-EMMA-0002) and $\emph{SPOCQ}$ (ANR-14-CE32-0019)  projects, the Conseil R\'egional PACA for the $\emph{Distance}$ project (Apex 2011: 2001-06938), the CNRS for the $\emph{Conneqt}$ project (PEPS INSIS 2011), the Majlis Amanah Rakyat (MARA), as well as  the European FP7-ITN for the $\emph{PICQUE}$ project (n.608062). The authors also acknowledge technical support from SCONTEL and IdQuantique and thank A. Martin, P. Vergyris and F. Kaiser for fruitful discussions.

\end{acknowledgements}




\end{document}